\documentclass[trackchanges,twocolumn]{aastex701}

\usepackage{amsmath}
\usepackage{enumerate}
\usepackage{graphicx}

\begin{document}

\title{Evidence for Multiple Orbiting Hotspots in the 340 GHz Variability of Sgr A*}
\correspondingauthor{Kazuki Yanagisawa}
\email{yanalgm0626@keio.jp}
\author[orcid=0009-0006-5208-3962,gname='Kazuki',sname='Yanagisawa']{Kazuki Yanagisawa}
\affiliation{School of Fundamental Science and Technology, Graduate School of Science and Technology, Keio University, 3-14-1 Hiyoshi, Kohoku-ku, Yokohama, Kanagawa 223-8522, Japan}
\email{yanalgm0626@keio.jp}

\author[orcid=0000-0002-5566-0634,gname='Tomoharu',sname='Oka']{Tomoharu Oka}
\affiliation{School of Fundamental Science and Technology, Graduate School of Science and Technology, Keio University, 3-14-1 Hiyoshi, Kohoku-ku, Yokohama, Kanagawa 223-8522, Japan}
\affiliation{Department of Physics, Institute of Science and Technology, Keio University, 3-14-1 Hiyoshi, Kohoku-ku, Yokohama, Kanagawa 223-8522, Japan}
\email{tomo@phys.keio.ac.jp}

\author[orcid=0009-0006-9842-4830,gname='Tatsuya',sname='Kotani']{Tatsuya Kotani}
\affiliation{School of Fundamental Science and Technology, Graduate School of Science and Technology, Keio University, 3-14-1 Hiyoshi, Kohoku-ku, Yokohama, Kanagawa 223-8522, Japan}
\email{sci.tatsu.729@keio.jp}

\author{Ryo Ariyama}
\affiliation{School of Fundamental Science and Technology, Graduate School of Science and Technology, Keio University, 3-14-1 Hiyoshi, Kohoku-ku, Yokohama, Kanagawa 223-8522, Japan}
\email{sirius@keio.jp}

\author{Kazuki Yanagihara}
\affiliation{School of Fundamental Science and Technology, Graduate School of Science and Technology, Keio University, 3-14-1 Hiyoshi, Kohoku-ku, Yokohama, Kanagawa 223-8522, Japan}
\email{lakalakalove.uver@keio.jp}

\author[orcid=0000-0002-9255-4742,gname='Yuhei',sname='Iwata']{Yuhei Iwata}
\affiliation{Mizusawa VLBI Observatory, National Astronomical Observatory of Japan, 2-12 Hoshigaoka, Mizusawa, Oshu, Iwate 023-0861, Japan}
\affiliation{Astronomical Science Program, Graduate Institute for Advanced Studies, SOKENDAI, 2-21-1 Osawa, Mitaka, Tokyo, 181-8588, Japan}
\email{yuhei.iwata@nao.ac.jp}
\begin{abstract}
We analyzed 11 epochs of archival Atacama Large Millimeter/submillimeter Array (ALMA) data to investigate flux density variability of Sgr A* at $340\,\mathrm{GHz}$. 
In one epoch, the light curve exhibits two short-timescale components with characteristic periods of $\sim$30 min and $\sim$50 min. While the corresponding peaks in the periodogram are highly significant under a white-noise assumption, their significance decreases below 3$\,\sigma$ when red-noise variability is taken into account, and we therefore do not regard them as statistically significant periodic detections. Nevertheless, the observed timescales are comparable to the orbital period near the innermost stable circular orbit of Sgr A*, and the light curve shows phase-dependent structure and amplitude evolution consistent with orbital modulation. 
We find that the variability is well described by a model involving multiple orbiting hotspots with decaying emission.
This interpretation suggests that both periodic and non-periodic variability in Sgr A* may arise from a common physical origin in orbiting structures within the accretion flow, providing a unified framework for its millimeter variability.
\end{abstract}


\section{Introduction} 
Sagittarius A* (Sgr A*) is a compact radio source located at the center of the Milky Way and is thought to host a supermassive black hole (SMBH) with a mass of $\sim 4\times10^6\,M_\odot$ \citep[e.g.,][]{abd2024improving}. 
Sgr A* provides an excellent laboratory for testing general relativity and studying accretion physics at event-horizon scales in the low-luminosity regime. In recent years, multiwavelength observations in the radio, mid-infrared, near-infrared, and X-ray bands have revealed its activity over wide wavelengths \citep[e.g.,][]{michail2021multiwavelength}.
Sgr A* is persistently bright in the radio band and can be readily detected at all times\citep[e.g.,][]{falcke1998simultaneous, subroweit2017submillimeter}.
In the near-infrared, it is continuously detected and exhibits both quiescent variability and occasional bright ``flares''.
During these events, significant flux variability is also observed. Flares have been detected in the radio band \citep[e.g.,][]{brown1982variability, witzel2021rapid}, the infrared band \citep[e.g.,][]{genzel2003near, eckart2004first, dodds2011two}, and the X-ray band \citep[e.g.,][]{baganoff2003chandra,yuan2015systematic, mossoux2020continuation}. 
These flares are characterized by large-amplitude flux variations on timescales ranging from several tens of minutes to a few hours \citep[e.g.,][]{genzel2003near,von2023general}.\\\indent
Transient periodic variabilities in Sgr A* have been reported across X-ray \citep[e.g.,][]{aschenbach2004x, belanger2006periodic}, infrared \citep[e.g.,][]{genzel2003near, michail2024multiwavelength}, and radio \citep{miyoshi2011oscillation}. 
The characteristic timescales of these reported variabilities are comparable to the orbital period near the innermost stable circular orbit (ISCO), $\sim$ 30 minutes. 
The ``hotspot'' model has been proposed as a plausible explanation for this variability. 
In this scenario, a localized emitting region orbiting around the SMBH produces flux variability through relativistic effects such as Doppler beaming and gravitational redshift.
Moreover, such hotspots are often associated with flare events that may be triggered by processes such as magnetic reconnection, making this model capable of providing a unified explanation for quasi-periodic variability that accompanies flares \citep[e.g.,][]{broderick2006imaging, trippe2007polarized, hamaus2009prospects, tiede2020spacetime}.\indent

Evidence for the presence of hotspots has been suggested through various observational approaches, including astrometric measurements, polarization observations, and flux variability analyses \citep[e.g.,][]{abuter2018detection,jimenez2020dynamically,wielgus2022orbital,abuter2023polarimetry}.
In particular, observations with the GRAVITY instrument have revealed orbital motion of compact emission regions through astrometric measurements, accompanied by variations in the linear polarization, providing strong evidence for orbiting hotspot scenarios.
In 230 GHz light curves obtained with the Atacama Large Millimeter/submillimeter Array (ALMA), \citet[I20+ hereafter]{iwata2020time} reported a transient flux variation with a timescale of $\sim$30 min. 
Using a same analysis method, \citet[Y25+ hereafter]{yanagisawa2025face} further identified a clear transient periodic variability with a period of $\sim$52 min at a significance level of 5$\,\sigma$. 
This variability was interpreted as arising from Doppler beaming produced by a hotspot orbiting on a nearly face-on accretion disk.\indent

However, these flare events are not persistent and are known to decay with time due to processes such as synchrotron cooling and adiabatic cooling associated with the expansion of the emitting region \citep[e.g.,][]{eckart2009modeling,dodds2009evidence,boyce2019simultaneous,michail2024multiwavelength,von2025first}. 
Considering these physical processes, the hotspots associated with flares are also expected to be transient structures with finite lifetimes.
Hotspots are expected to be short-lived, with lifetimes comparable to or shorter than their orbital periods\citep[e.g.,][]{yfantis2024fitting,yfantis2024hot}. As a result, periodic variability may not persist over multiple cycles, making it difficult to detect clear periodic signatures in the observed light curves.
The reported periodic or quasi-periodic variability is not persistent and often appears only temporarily. Although the origin of such sporadic periodicity remains unclear, it may provide important clues about the geometry, kinematics, and magnetic field structure of the innermost accretion flow.
\section{Observations and Data Reduction} \label{observation}
We systematically surveyed 340 GHz data of Sgr A$^*$ in the ALMA Science Archive.  The selection criteria required (1) focusing on Sgr A$^*$'s field of view, (2) frequency band 7, and (3) integration time exceeding 1 hour.  
We identified five ALMA projects conducted between 2016 and 2021 that satisfy these criteria. None of these observations included polarization measurements. We applied self-calibration and the CLEAN algorithm \citep{hogbom1974synthesis} to these data using the Common Astronomy Software Applications (CASA) package \citep{mcmullin2007casa, bean2022casa}. 
For the raw ALMA data, standard calibration scripts provided by the observatory (ScriptForPI.py) were applied to perform flux, phase, and bandpass calibration. 
After this standard calibration, we performed iterative phase self-calibration on Sgr A*, following the procedures described in I20+ and Y25+.

\section{Variability analysis} \label{results}
\begin{figure}[t]
\begin{center}
\includegraphics[width=0.45\textwidth]{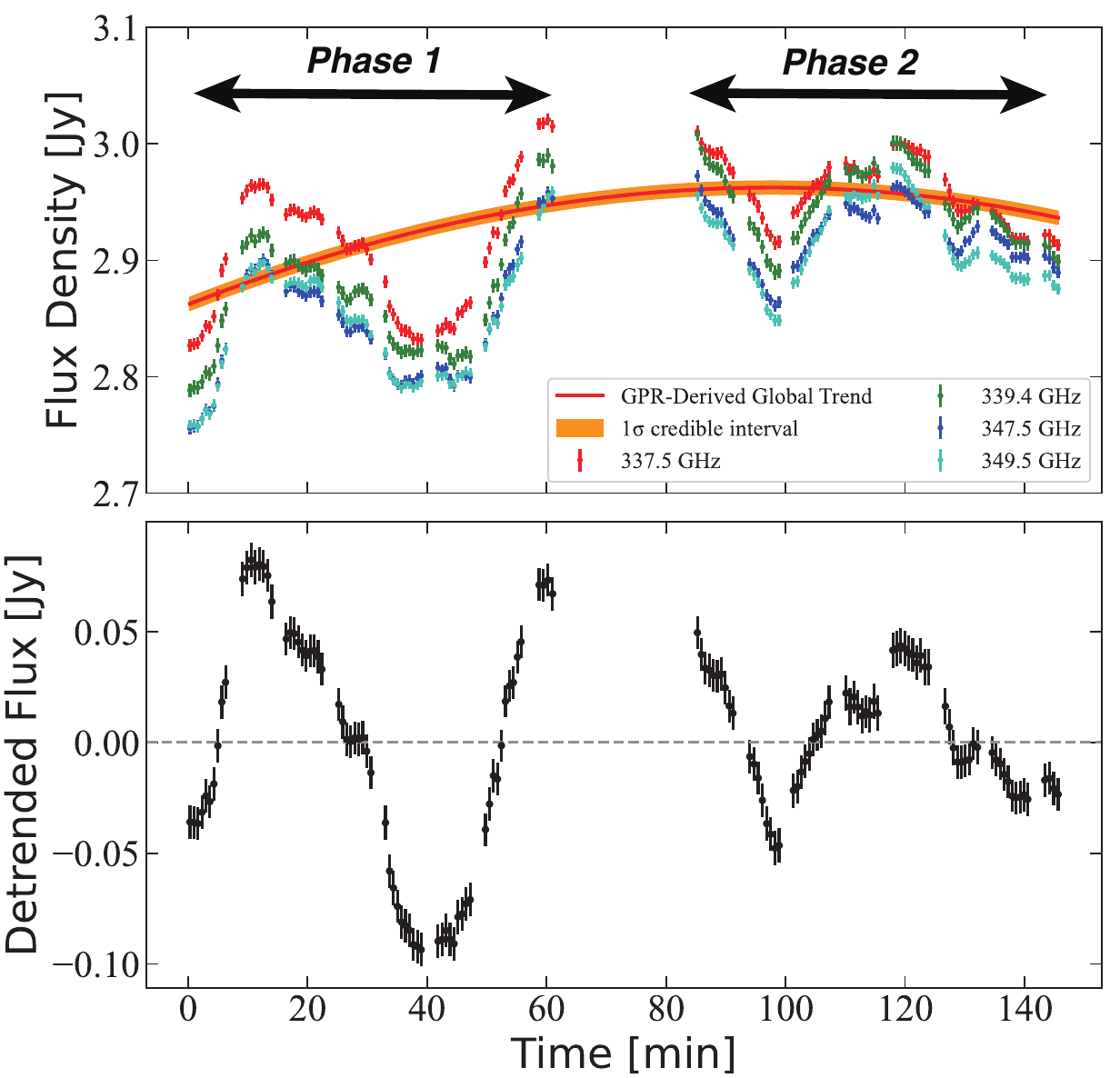}
\caption{{\it Top}: Light curve of Sgr A$^*$ observed on 2016 August 31 with ALMA at 337.5 GHz (red), 339.4 GHz (green), 347.5 GHz (blue), and 349.5 GHz (cyan). The horizontal axis shows time in minutes relative to the observation start. The light curve is divided into two intervals: $t=0\text{--}60\,\mathrm{min}$ (Phase 1) and $t=80\text{--}140\,\mathrm{min}$ (Phase 2). The red curve shows the inferred trend, with the shaded region indicating the 1$\sigma$ uncertainty.
{\it Bottom}: Light curve at 337.5 GHz after subtracting the long-timescale variability. The error bars represent the uncertainties propagated from both the observational errors and the GPR prediction.}
\label{01080}
\end{center}
\end{figure}

\begin{figure}[t]
\begin{center}
\includegraphics[width=0.47\textwidth,trim=0cm 0cm 1cm 0cm,clip]{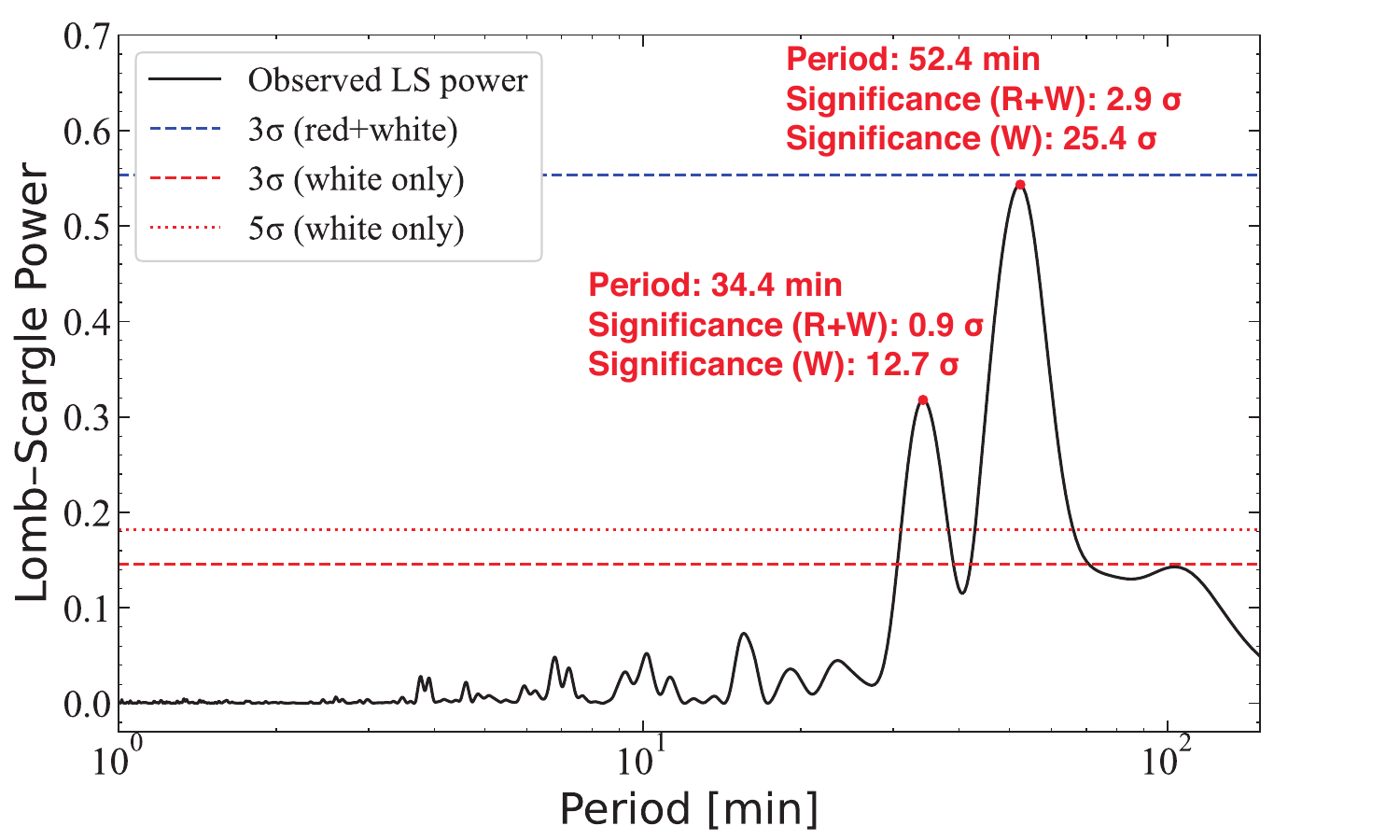}
\caption{Generalized Lomb-Scargle periodogram of the light curve observed on 2016 August 31. The black line shows the LS power of the observed light curve in this study. 
FAP is estimated by generating $10^5$ simulated light curves for (1) red noise plus white noise (R+W) and (2) white noise only (W). For the R+W case, the 3$\,\sigma$ significance level is shown by the blue dashed line. For the W case, the 3$\,\sigma$ and 5$\,\sigma$ significance levels are indicated by the red dashed and dotted lines. 
Peaks in the LS power are marked with red points, with the corresponding FAP values indicated.}
\label{GLS}
\end{center}
\end{figure}

In total, we obtained light curves for 11 epochs, each with observation durations ranging from 35 to 150 minutes. All data were binned with a time resolution of $\sim$ 1 minute. 
The total observing time amounts to about 14 hours. Among these observations, we focus on the light curve obtained on 2016 August 31 during ALMA Cycle 3 (project 2015.1.01080.S; PI: M. Tsuboi), which spans a continuous 150-minute observing period and exhibits clear short-timescale variability (Figure \ref{01080} top).
This light curve exhibits several characteristic features: (1) variability with a single periodic component in phase 1 ($t = 0\text{--}60\,\mathrm{min}$), (2) variability with more complex structures in phase 2 ($t = 80\text{--}140\,\mathrm{min}$), (3) a time-dependent amplitude, and (4) modulation by a longer-timescale variation exceeding the observation duration. 
These features are consistently observed in all four analyzed spectral windows (SpWs).\\\indent
The observed flux exhibits a gradual variation on timescales longer than the observing window, upon which shorter quasi-periodic structures are superposed.
Since our primary interest is in modeling short-timescale variability, it is necessary to separate these components from the global flux variation.
To isolate these variations, we employed Gaussian Process Regression (GPR) and subtracted the long-timescale variability inferred from the observed light curve. 
The correlation function was modeled using a radial basis function (RBF) kernel \citep{williams2006gaussian}, as it provides a smooth representation of long-timescale variability and avoids introducing spurious small-scale structure into the detrended light curve.
The correlation length in RBF kernel was fixed to 300 min, corresponding to the average of the characteristic variability timescales reported in three previous studies \citep{meyer2006near,dexter20148,wielgus2022orbital}.
Measurement uncertainties are incorporated as Gaussian white noise. Using this kernel, we construct the full covariance matrix and infer the large-scale variability component.
The detrended light curve at 337.5 GHz obtained by subtracting the predicted offset is shown in Figure \ref{01080} (bottom). For comparison, we also tested a Matérn kernel with similar characteristic length scales. Although the Matérn kernel yields a less smooth trend, the resulting detrended light curves and model fits are qualitatively consistent. This indicates that our results are not sensitive to the choice of kernel. 
We verified that applying the same detrending procedure to other epochs does not introduce artificial periodic features.\\\indent
To search for periodic signals, we applied the Generalized Lomb-Scargle (GLS) periodogram \citep{zechmeister2009generalised} to the light curves. 
The false-alarm probability (FAP) for each peak in the periodogram is estimated using $10^5$ simulated light curves that incorporate both red noise, assuming a power spectral density of PSD$\propto f^{-2}$, and Gaussian white noise consistent with the observational uncertainties (Figure \ref{GLS}).
We adopt a false alarm probability (FAP) of $< 1.35 \times 10^{-3}$,corresponding to a 3$\,\sigma$ significance level,as the criterion for significant periodicity.
As a result, for the light curve obtained on 31 August 2016, the periodogram exhibits two peaks with a significance close to the 3$\,\sigma$ level.
When only white noise is considered, peaks at $\sim$52 min and $\sim$34 min are obtained with significances of $\sim25.4\,\sigma$ and $\sim12.7\,\sigma$. 
When both white and red noise are included, however, these are reduced to $\sim2.9\,\sigma$ and $\sim0.9\,\sigma$. We therefore do not regard these periodicities as statistically significant detections. 
Nevertheless, the corresponding timescales are comparable to the orbital period near the ISCO of Sgr A*, and the light curve exhibits phase-dependent structure and amplitude evolution suggestive of orbital modulation. 
Taken together, these properties motivate an interpretation in terms of relativistic Doppler beaming from orbiting hotspots, as explored below.
\section{Multiple Hotspot Model}
\begin{figure*}[t]
\begin{center}
\includegraphics[width=\textwidth,trim=0cm 3.5cm 0cm 0cm,clip]{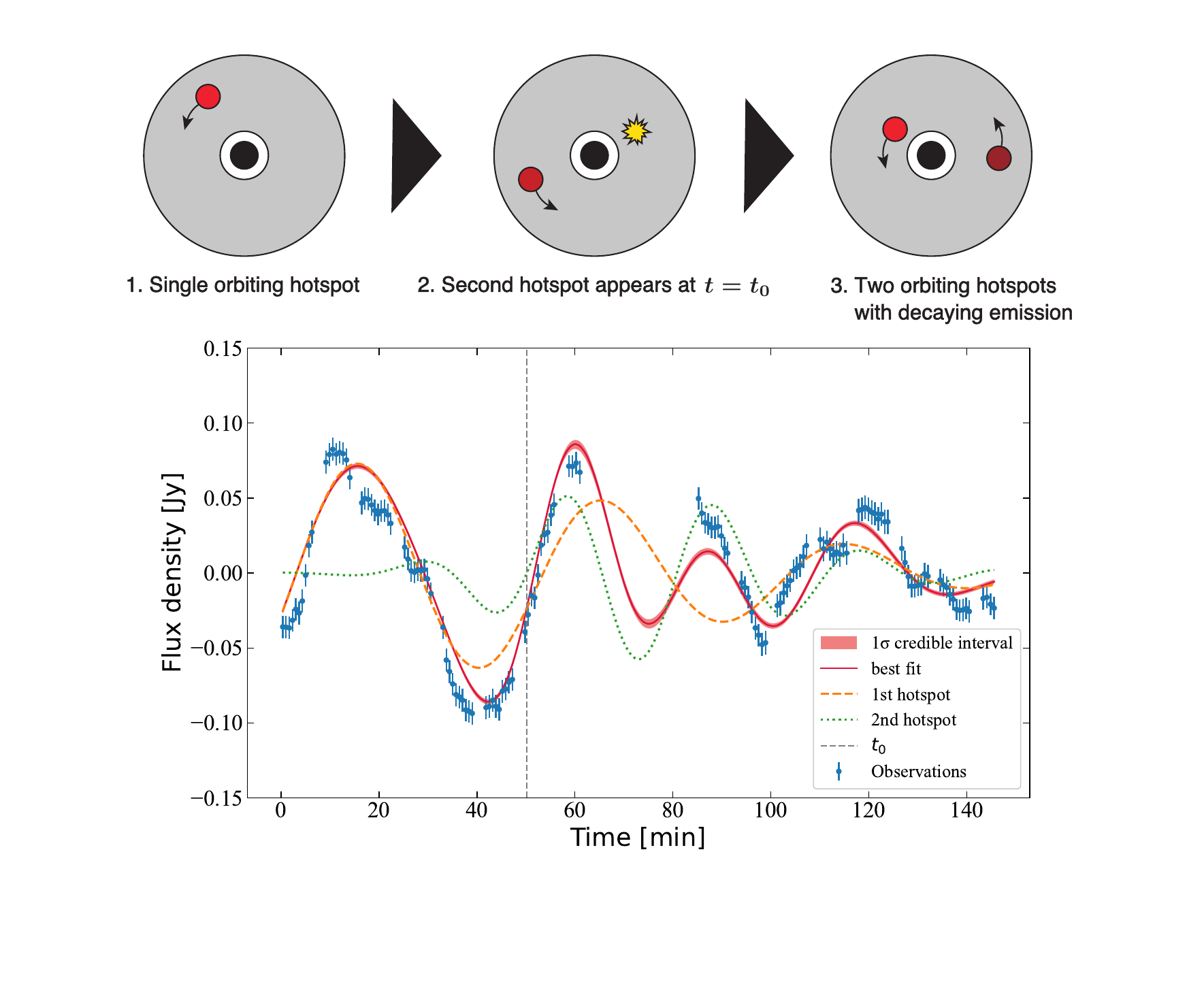}
\caption{{\it Top}: Schematic diagram of the multiple hotspot model. Initially, a single hotspot revolves around the black hole. A second hotspot appears at $t=t_0$, after which both orbit while gradually fading in energy. The color variation reflects this energy loss.
{\it Bottom}: Light curve fitted using the MCMC method. The blue points represent the observed data at $337.5\,\mathrm{GHz}$, and the red curve shows the best-fit model given by Eq. (\ref{eq:hotspot_1}). The shaded region around the model curve indicates the $1\,\sigma$ credible interval. 
The orange and green dashed lines indicate the contributions from the 1st and 2nd hotspots, respectively. The gray dashed line marks $t_0$. 
The bump component around $t\sim90$ min, which is not reproduced by the model, may arise from effects not included in the present model, such as relativistic effects.}
\label{hotspot}
\end{center}
\end{figure*}
In addition to the presence of two peaks at characteristic timescales, the observed decay in their amplitudes motivates us to adopt a multiple-hotspot model.
In this model, a single hotspot orbits during phase 1, after which a second hotspot emerges. Subsequently, the two hotspots continue orbiting while gradually losing energy (Figure \ref{hotspot} top).
Within this framework, the flux density of Sgr A* can be written as
\begin{multline}
  F_{\rm obs}(t)=A_1(t)\sin{\Bigl(\frac{2\pi}{T_1}t-\phi\Bigr)}\\
  +U(t-t_0)A_2(t-t_0)\sin{\Bigl[\frac{2\pi}{T_2}(t-t_0)\Bigr]}.
\label{eq:hotspot_1}
\end{multline}
Here, $\phi$ represents the phase offset, $T_1$ and $T_2$ are the orbital periods, and $t_0$ denotes the time at which the second hotspot emerges.
The time-dependent amplitudes $A_i(t)$ ($i=1,2$) represent the decaying amplitudes and are assumed to follow
\begin{equation}
  A_i(t)=A_{i0}\exp{\Bigl(-\frac{t^2}{2\sigma_i^2}\Bigr)},
\label{eq:hotspot_2}
\end{equation}
where $\sigma_i$ is the decay timescale, treated here as a phenomenological parameter. We adopt a Gaussian function as a simple phenomenological model for the intrinsic evolution of each hotspot, capturing a single characteristic timescale with minimal free parameters\citep[e.g.,][]{von2025first}. We note that this choice is not unique, and the inferred parameters may depend on the assumed functional form.
The function $U(t)$ is a sigmoid function that represents the emergence of the second hotspot during the observation,
\begin{equation}
  U(t)=\frac{1}{1+\exp(-t/\tau_{\rm occ})},
\label{eq:hotspot_3}
\end{equation}
where $\tau_{\rm occ}$ is the characteristic timescale for the hotspot appearance.
Although the variability is modeled here as sinusoidal, this assumption is motivated by the Doppler beaming model proposed by Y25+, in which the flux variation from an orbiting hotspot can be approximated by a sinusoidal function in face-on orbits.
Eq. (\ref{eq:hotspot_1}) was fitted to the light curve shown in Fig. \ref{01080} (bottom) using the Markov Chain Monte Carlo (MCMC) method. The resulting best-fit model and parameters shown in Figure \ref{hotspot} (bottom) and Tabel \ref{table_fit}.
The two periods obtained from the fit are $T_1=50.4\pm0.3\,\mathrm{min}$ and $T_2=30.6\pm0.3\,\mathrm{min}$, which are consistent with previously reported timescales \citep[e.g., I20+;][Y25+]{abuter2023polarimetry,michail2024multiwavelength}. Furthermore, the decay timescales of the amplitudes were found to be $\sigma_1=70.3\pm2.5\,\mathrm{min}$ and $37.7^{+1.6}_{-1.5}\,\mathrm{min}$.
We note, however, that these values should be regarded as approximate constraints, as the limited duration of the light curve does not allow us to robustly probe variability on significantly longer timescales.

\begin{table*}[htbp]
   \centering
  \caption{Parameters obtained from the fitting}
  \label{table_fit}
  \begin{tabular}{ccccccc}
    \hline
      & $A_{i0}$ [Jy] &$T_i$ [min]& $\sigma_i$ [min]&$\phi$ &$t_0$ [min]&$\tau_{\rm dec}$ [min]\\
    \hline \hline
     1st hotspot &$(7.5\pm0.2)\times10^{-2}$&$50.4\pm0.3$&$70.3\pm2.5$&$0.39\pm0.03$&--&--\\
     2nd hotspot&$(7.8\pm0.4)\times10^{-2}$&$30.6\pm0.3$&$37.7^{+1.6}_{-1.5}$&--&$50.2\pm0.3$&$10.9^{+1.0}_{-0.9}$ \\
     \hline
  \end{tabular}
\end{table*}

\section{Discussion}\label{dis}
\subsection{Cooling of the Hotspots}
In this study, the amplitude of the flux variability is found to decay on a timescale of several tens of minutes. 
One possible origin of this decay is synchrotron cooling, as suggested in previous studies \citep[e.g.,][]{eckart2009modeling,dodds2009evidence,von2025first}.
We consider the synchrotron cooling timescale $t_{\mathrm{syn}}$, which is given by
\begin{equation}
  t_{\rm syn}=\frac{6\pi m_{\rm e}}{\sigma_TB^2\gamma},
\label{syncro_1}
\end{equation}
where $m_{\rm e}$ is the electron mass, $B$ is the magnetic field strength, $\sigma_T$ is the Thomson cross section, and $\gamma$ is the Lorentz factor of the emitting electrons.
To estimate $\gamma$, we use the relation for the synchrotron frequency
\begin{equation}
\nu_{\rm syn}=\frac{3}{2}\gamma^2\frac{eB}{2\pi m_{\rm e}c}\sin{\theta},
\label{syncro_2}
\end{equation}
where $e$ is the elementary charge, $c$ is the speed of light, and $\theta$ is the pitch angle.
Assuming $\nu_{\mathrm{syn}} = 337.5\,\mathrm{GHz}$ and $\sin\theta = 1$, and adopting a magnetic field strength of $B = 44\,\mathrm{G}$ at an orbital radius of $\sim 3.1\,r_{\rm s}$ \citep{von2025first}, we obtain a synchrotron cooling timescale of $t_{\mathrm{syn}} \sim 150\,\mathrm{min}$. 
The derived timescale is formally independent of whether the electron population is thermal or non-thermal, as it is defined for a given Lorentz factor, although the characteristic value of $\gamma$ depends on the underlying electron distribution.
This value is longer than the decay timescales of the amplitudes derived in this study. 
However, when focusing on the variation of the spectral index $\alpha$ ($f\propto\nu^{-\alpha}$), we find that the $\alpha$ decreases by $\sim\,0.4$ as the amplitude decays. 
The observed decrease in $\alpha$ therefore suggests that synchrotron cooling alone cannot account for the observed decay. This implies that the decay timescale $\sigma$ may reflect the combined effects of multiple physical processes. We note, however, that alternative interpretations, such as a superposition of multiple emission components, are also possible; nevertheless, the quasi-periodic variability supports the hotspot scenario in this case.\\\indent
Another possible cooling mechanism is the adiabatic expansion of the hotspot \citep[e.g.,][]{eckart2006flare,subroweit2017submillimeter}. 
\citet{michail2024multiwavelength} reported that, for a magnetic field strength of $50\,\mathrm{G}$, the adiabatic expansion timescale of a hotspot is on the order of $\sim1\,\mathrm{hr}$, which is broadly consistent with the timescales derived in this study. 
The observed change in the spectral index could be explained by variations in optical depth associated with the adiabatic expansion.

However, the decay timescale alone is insufficient to uniquely identify the physical mechanism responsible for the variability. 
To test the adiabatic expansion scenario more robustly, additional observational constraints are required, such as polarization properties or simultaneous multi-frequency time delays expected from an expanding synchrotron-emitting blob. 
Therefore, future multi-band observations will be essential to further investigate this possibility.
\subsection{Non-periodic variabilities Scenario in Sgr A*}
Although periodic variability is occasionally detected in Sgr A*, its flux variability is predominantly non-periodic and lacks clear periodic signatures during most observations (I20+; Y25+). 
Various models have been proposed to explain the periodic variability, including orbiting hotspot scenarios. 
In contrast, non-periodic variability has often been attributed to stochastic fluctuations in physical quantities such as magnetic field strength, density, magnetisation, and temperature in turbulent accretion flows\citep{dexter20148,chatterjee2021general, murchikova2021second}. 
These phenomena have generally been treated as distinct processes depending on their characteristic timescales.\\\indent
In this study, we find evidence that two hotspots are simultaneously orbiting in the accretion flow. 
This result suggests that multiple hotspots, rather than a single hotspot, can coexist and rotate simultaneously around Sgr A*. 
On the other hand, in our dataset, only one out of 11 light-curve segments shows variability consistent with orbital motion, while the remaining segments do not exhibit clear periodic signatures. 
This provides an important observational constraint, suggesting that such periodic variability is either intrinsically rare or only intermittently detectable in the submillimeter regime.
Based on this observational constraint, we propose a scenario in which non-periodic variability can be explained within a multiple-hotspot model.
Under typical conditions, several hotspots may exist in the accretion disk. Each hotspot orbits with a finite lifetime, while new hotspots are continuously generated. As a result, the individual periodic signals do not build up sufficient coherence over the limited duration of the observations, making a clear periodic signature difficult to detect.
On the other hand, if the number of hotspots decreases due to a transient event such as a flare, a single hotspot may become dominant, allowing periodic variability to emerge more clearly.
Indeed, previous studies have suggested a strong connection between flares and the appearance of periodic variability \citep[e.g.,][]{genzel2003near,hamaus2009prospects,abuter2018detection,wielgus2022orbital}. 
This scenario may therefore provide a unified interpretation of both periodic and non-periodic variability in Sgr A* within the hotspot model.

\section{Summary}
In this work, we analyzed 11 epochs of ALMA archival data to investigate flux variability in Sgr A$^*$.
In one of the light curves, two peaks at characteristic timescales are identified. While these features are not statistically significant as periodic signals in the presence of red noise, the observed variability also exhibits a time-dependent decay in amplitude, which motivates an interpretation in terms of a multiple-hotspot model with orbiting components undergoing energy loss. 
Within this framework, the inferred orbital periods are $\sim$30 min and $\sim$50 min, with corresponding amplitude decay timescales of $\sim$70 min and $\sim$37 min. 
These timescales may be associated with synchrotron cooling, adiabatic expansion of the hotspots, or a combination of both, although further constraints require simultaneous multiwavelength observations.
The presence of multiple orbiting hotspots suggests that both periodic and aperiodic variability in Sgr A* can be described within a unified hotspot framework.\\\indent
This study demonstrates that millimeter-wavelength flux variability can serve as a powerful probe of the physical conditions in the immediate vicinity of SMBHs. By linking ALMA variability observations with X-ray and near-infrared measurements, our results provide a bridge toward understanding the short-timescale variability processes in Sgr A*.
Furthermore, simultaneous analysis of linear polarization variability may enable us to disentangle variability induced by the orbital motion of the hotspot from intrinsic variations in its emission.
Future long-term and high-cadence monitoring of Sgr A* with ALMA, combined with simultaneous multi-wavelength observations, will enable further investigation of hotspot dynamics and may lead to a unified understanding of the flux variability processes in Sgr A*.

\begin{acknowledgments}
This paper makes use of the following ALMA data: ADS/JAO.ALMA\#2015.1.01080.S, \#2016.1.00517.S, \#2016.1.01430.S, \#2018.A.00050.T, and \#2019.1.01559.S.
ALMA is a partnership of ESO (representing its member states), NSF (USA), NINS (Japan), NRC (Canada), NSTC and ASIAA (Taiwan), and KASI (Republic of Korea), in cooperation with the Republic of Chile. The Joint ALMA Observatory is operated by ESO, AUI/NRAO, and NAOJ.
T.O. acknowledges the support from JSPS Grant-in-Aid for Scientific Research (A) No.20H00178.
\end{acknowledgments}


\facility{ALMA}

\software{CASA (version 4.7, 4.7.2, 5.4.0-68, 5.4.0-70, 6.2.1.7)}
\newpage
\bibliography{refer_yana}
\bibliographystyle{aasjournalv7}

\end{document}